# Time-Frequency Trade-offs for Audio Source Separation with Binary Masks

Andrew J.R. Simpson [#1]

[#] *Centre for Vision, Speech and Signal Processing, University of Surrey*
*Guildford, UK*
[1]`andrew.simpson@surrey.ac.uk`

*Abstract*— **The short-time Fourier transform (STFT) provides the foundation of binary-mask based audio source separation approaches. In computing a spectrogram, the STFT window size parameterizes the trade-off between time and frequency resolution. However, it is not yet known how this parameter affects the operation of the binary mask in terms of separation quality for real-world signals such as speech or music. Here, we demonstrate that the trade-off between time and frequency in the STFT, used to perform ideal binary mask separation, depends upon the types of source that are to be separated. In particular, we demonstrate that different window sizes are optimal for separating different combinations of speech and musical signals. Our findings have broad implications for machine audition and machine learning in general.**

## I. INTRODUCTION

For many years, audio source separation methods, and machine listening methods in general, have been based on spectrograms computed using the short-time Fourier transform (STFT) [1]-[10]. However, despite the ubiquitous application of, and dependence upon, the STFT, little consideration appears to be given to the trade-off between time and frequency resolution which is defined by the STFT window size; window size is chosen arbitrarily (or by convention) for most applications and once chosen is not considered further. However, given that audio signals exist on a continuum between steady sinusoids and time-varying stochastic signals (noise), a one-size-fits-all approach does not make much sense. In particular, using a single, arbitrary and fixed spectrogram representation to separate diverse signals such as music and speech would seem less than ideal. In this paper, we explore the time-frequency trade-off for ideal binary mask separation applied to different signals.

The ideal binary mask is constructed by assigning each element of a mixture spectrogram to that source whose individual spectrogram features the greatest magnitude corresponding to that element. This binary assignment (or classification) results in a mask which may be multiplied with the mixture spectrogram before inverting the masked spectrogram to resolve the separated components into estimates of the source audio signals.

A particular issue with the binary mask separation approach is that the superposition of signals (i.e., within a single component of the spectrogram) is not well handled. Since the time and frequency resolution to some extent defines the degree to which such overlap of energy occurs, for a given spectrogram representing a mixture of signals, it is possible to optimize the trade-off in an informed way. In particular, we may adjust the window size of the STFT so that we maximize separation quality for a given source or combination of sources. We may even, in principle, extract different sources using different window sizes which maximize the mapping of source signal statistics to the distribution of resolution for a given spectrogram representation. Hence, by manipulating the STFT window size, we may mitigate the problem of superposition or overlap in an informed way.

In this paper, we construct ideal binary masks to separate speech and music signals featuring contrasting signal characteristics. Using objective source separation quality measures [11], we show that each source and each combination of sources (mixture) has a distinct optimal STFT window size for separation.

## II. METHOD

We consider the problem of fully-informed separation of linear mixtures of pairs of typical speech and music sources. Two brief speech excerpts were chosen (a male and female voice). One excerpt of single piano notes and one excerpt of single snare drum hits were also chosen. The excerpts were of 4s duration and were sampled at 44.1 kHz. The various audio signals were transformed into spectrograms using the STFT with window sizes between 2 and 2^14 samples, with 1-sample step size (i.e., with maximum overlap) and a Hanning window, giving spectrograms with various numbers of frequency bins. From the source spectrograms a binary mask was constructed by element-wise comparison of the magnitudes and assignment of a '1' to any element where the designated 'target' source had greater magnitude. By multiplying this binary mask with the mixture spectrogram a masked estimate of the target source was obtained. The masked estimate of the respective alternate source was obtained by multiplying the mixture spectrogram with an element-wise subtraction of the binary-mask from 1.

This procedure was carried out for mixtures of piano/drum, drum/voice and male/female voice at various STFT window sizes (range: $2...2^{14}$). The resulting masked spectrograms were inverted with a standard overlap-and-add procedure. Separation quality was then measured using the BSS-EVAL toolbox [11] and is quantified in terms of signal-to-distortion ratio (SDR), signal-to-interference ratio (SIR) and signal-to-artefact ratio (SAR), computed by comparing the estimated sources with the original sources.

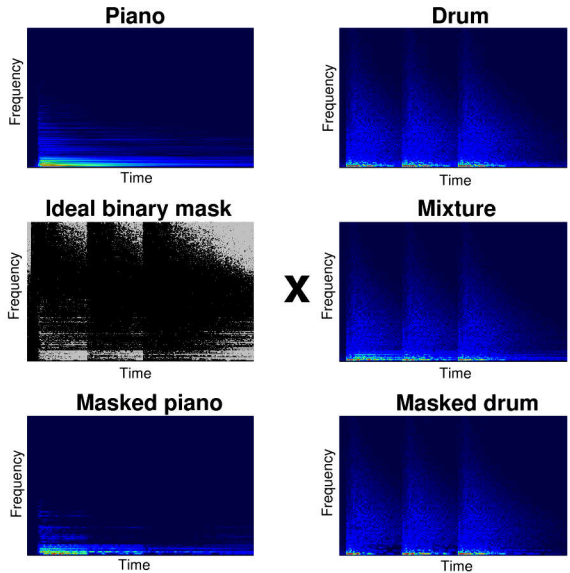

**Fig. 1. Illustration of ideal binary mask: piano versus drum.** The upper pair of spectrograms plot a ~2-second excerpt from the piano and drum sources (featuring a single piano note and several drum hits), illustrating the tonal energy of the piano as compared to the noisy energy of the (snare) drum. The middle spectrograms plot the ideal binary mask and the mixture. The lower spectrograms plot the sources extracted from the mixture using the binary mask (and the inverse binary mask) repsectively. Window size for this example was 512 samples. Note the frequency axis represents the range 0 – 22 kHz on a linear axis.

### III. RESULTS

Fig. 1 plots spectrograms illustrating the stages of ideal binary mask mixture and separation for a portion (~2 seconds) of the piano/drum case. The spectrograms for the piano and drum sources are shown at the top. The middle panel plots the ideal binary mask and the mixture spectrogram. At the bottom of Fig. 1 are plotted the respective spectrograms representing the audio separated using the binary mask. The piano and drum provide a compelling example because the piano (top left of Fig. 1) produces mainly tonal energy whilst the drum (top right of Fig. 1) produces mainly noise energy. As a result, the binary mask has a noisy component which corresponds to the energetic fluctuations in the drum hits, which cause 'noisy' modulations of the mask during the higher energy (earlier) portion of the piano note.

Fig. 2 plots the various source separation quality measures computed using the BSS-EVAL toolkit [11]. Fig. 2a plots the separation quality measures for the example (as plotted in Fig. 1) of the drum and piano mixture. In general, extraction of the drum is more successful. This is presumably the result of there being more energy in the drum signal and, hence, the drum dominates the binary mask. This is reflected in the artefacts (SAR, middle) and overall distortion (SDR, right). Furthermore, performance generally improves with increasing window size, suggesting the importance of frequency selectivity for the piano. Hence, we may conclude that to separate drum/piano mixtures, maximum window size is preferable due to the steady, tonal energy of the piano.

Fig. 2b plots the same comparisons for the drum/voice mixture. In this case, the drum appears to dominate overall by virtue of its energy profile. However, in this case the functions feature clear band-pass characteristics which illustrate that the optimal window size for separating drum and voice combinations is more tightly tuned to around 2-4,000 samples.

Fig. 2c plots the same comparisons for the male/female voice mixture. As with the voice/drum mixture, the functions feature clear band-pass characteristics. Interestingly, the peaks do not occur at the same window size; for the male voice, peak SIR occurs at a window size of around 1,000 samples, whereas for the female voice peak SIR occurs for a window size of around 3-5,000 samples. Thus, we may conclude that (from this brief example) window size is capable of discriminating between the two voices and that optimum extraction of each voice (from the same mixture) requires a different window size. In both Fig. 2b and 2c, the band-pass nature of the functions may be partly due to the fact that speech features a combination of tonal and noisy components. This combination implies that separation quality depends on a trade-off between the tonal component and the noisy component. This is in contrast to the more stereotypical tonal piano and noisy drum, whose separation results are more straight forward in terms of interpretation.

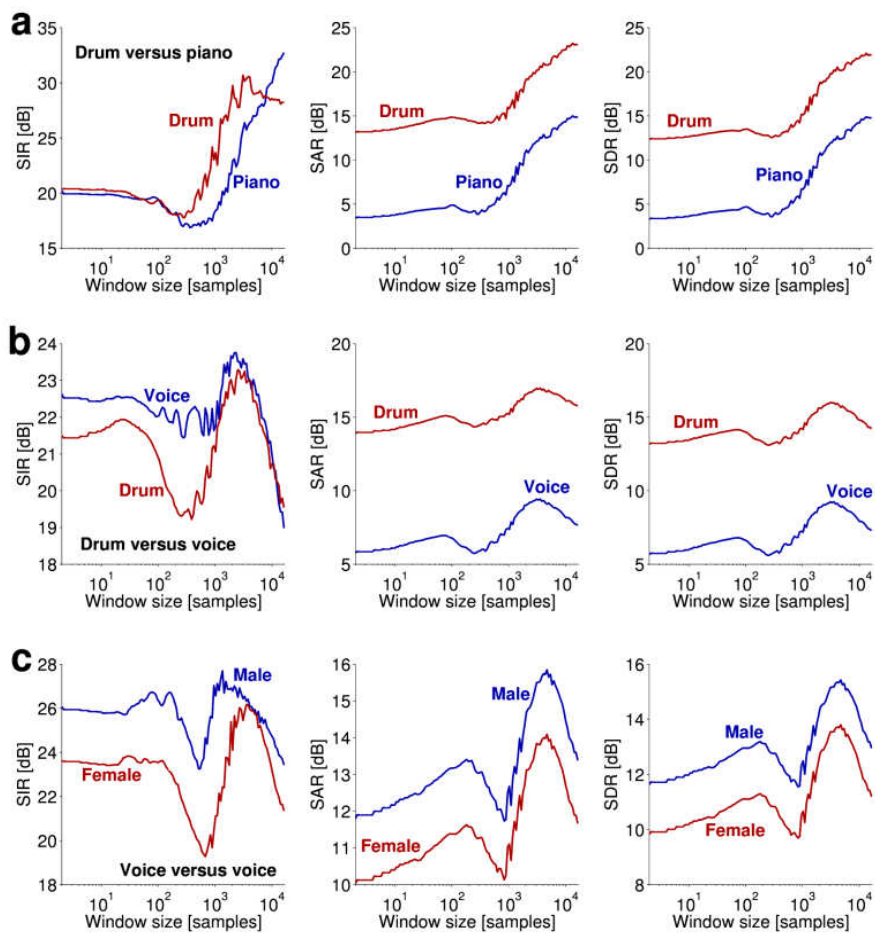

**Fig. 2. Separation quality as a function of STFT window size.** For the various combinations of drum/piano/voice, ideal binary mask source separation quality is evaluated in terms of signal-to-distortion ratio (SDR), signal-to-interference (SIR), signal-to-artefact ratio (SAR), computed from using the BSS-EVAL toolkit [11]. **a** plots the results for the mixture of drum and piano, **b** plots the same for the mixture of drum and voice, and **c** plots the same for the mixture of male/female voices.

IV. DISCUSSION AND CONCLUSION

We have demonstrated that STFT window size is critical in the application of binary-mask based source separation. We have demonstrted that separation quality may be optimized for a given source by adjustment of the window size. We have also demonstrated that one window size does not fit all. Our results suggest that broad classes of source (voices, percussion, tonal/non-tonal, etc), and even the various combinations of different classes of source, may require different window sizes for optimal separation using spectrogram based approaches.

These simple observations have relatively broad implications for machine listening and machine learning where it is applied to spectrogram representations. In particular, where spectrogram-based methods are evaluated, it is possible that results are confounded by interactions between signals and spectrogram window size. More generally, it would appear that any machine learning approach to source separation would likely benefit from at least careful consideration of the time-frequency trade-off. In principle, such approaches would likely benefit from tailored window-sizing on a per source or per class basis. For example, window size might be matched to a class of sound source according to some statistical measures of its energy profile.

ACKNOWLEDGMENT

AJRS was supported by grant EP/L027119/1 from the UK Engineering and Physical Sciences Research Council (EPSRC). Data and materials are available from the author on request.